
\overfullrule0pt

 \newtoks\slashfraction
 \slashfraction={.13}
 \def\slash#1{\setbox0\hbox{$ #1 $}
 \setbox0\hbox to \the\slashfraction\wd0{\hss \box0}/\box0 }

  \def\Buildrel#1\under#2{\mathrel{\mathop{#2}\limits_{#1}}}

\def\lozenge{\boxit{\hbox to 1.5pt{%
             \vrule height 1pt width 0pt \hfill}}}

\def\usp{{\underline p}}

\def\H{{\cal H}}
\def\K{{\cal K}}
\def\L{{\cal L}}
\def\M{{\cal M}}
\def\N{{\cal N}}
\def\P{{\cal P}}


 \doublespace
 \pubnum{6391}
\date{August 1993}
 \pubtype{T/E}
 \titlepage
\vfill
 \title{Parity and Front-Form Quantization of Field Theories
 \doeack}
 \author{ Ovid C. Jacob\foot{jacob@slacvm.slac.stanford.edu}
 }
 \SLAC
%
%
\abstract
Recently, we
 proposed   a new front-form quantization
which treated both the $x^{+}$ and the $x^{-}$  coordinates as
          front-form   'times.' This
quantization  was found to preserve parity explicitly.
In this paper we extend this construction to fermion fields in the
context of the Yukawa theory. We quantize        this theory using
a method     proposed originally by Faddeev and Jackiw .
We find that $P^-$ {\it and} $P^+$ become dynamical and that the
theory is manifestly invariant under parity.
\vfill
\centerline{Submitted to Phys. Lett. B}
\chapter{Introduction}
Front-form quantization is usually done by quantization along
the front $x^+=const  $ , though this causes some uneasiness
\REF\ji{X. Ji, Comm. in Part. and Nucl. Phys. {\us 21}, 123 (1993)}
\refend .
Usually this is done by quantizing a system with constraints
\REF\cd{Kurt Sundermeyer, 'Constrained Dynamics', Springer-Verlag,
New York, 1982 }
\refend
\REF\gt{D. M. Gitman and I. V. Tyutin, 'Quantization of Fields with
Constraints', Springer-Verlag, New York, 1990}
\refend
\REF\gov{Jan Govaerts , 'Hamiltonian Quantization and Constrained
Dynamics', Leuven University Press, Leuven, Belgium, 1991}
\refend
\REF\ht{Marc Henneaux and Claudio Teitelboim, 'Quantization of Gauge
Systems', Princeton University Press, Princeton, NJ, 1992}
\refend .
In a previous paper
\REF\pclcq{Ovid C. Jacob, 'Parity Conserving Light-Cone Quantization',
SLAC-PUB-6188,May 1993, hep-th/9305076, Submitted to Nucl, Phys. B.}
\refend ,
we   introduced a quantization which treated $x^+$ and
$x^-$  on equal footing. The main argument given was that
this new approach was  manifestly parity invariant.
We also pointed out that this new approach had the same number
of degrees of freedom as the equal-time approach. For the
scalar case, the second order differential equation was specified
by the two boundary conditions in both the equal time, as well as
  in the front-form case, where we took both
$x^+$ and $x^-$ as front-form 'times'
\REF\steinh{See also the Introduction in
Paul Joseph Steinhardt, Ann. Phys. {\us 128}, 425 (1980) where he
also points out that one needs to initialize the data along both
characteristics. He does not follow this point though, and proceeds
to study the nature of constraints, proper and improper , and their
mathematically consistent treatment.
Our approach is different from his in that we treat
$x^+$ and $x^-$ on equal footing while he does not. Also, there are
no constraints here  .}
\refend .

We want to present here a new argument, due to Robertson and
McCartor
\REF\rm{Gary McCartor and David Robertson,
Zeit. f. Phys. \us{C53}, 679, (1992)}
\refend  ,
which we feel is even more compelling. For this we will consider
the Yukawa model with scalar  mass $\mu$  and fermion mass $m$
in $3+1$-dimensions:
$$\L = {1\over 2}(\partial_\nu \phi \partial^\nu \phi - \mu^2 \phi^2)
+ {i\over 2} \partial^{\nu}\overline{\psi} \gamma_{\nu} \psi
- \overline{\psi} {i\over 2} \partial^{\nu}\gamma_{\nu} \psi
 -m \overline{\psi} \psi - g \overline{\psi} \psi \phi   \eqn\a $$

The equations of motions for the fermion  field are
$$ i \partial^- \psi_+ = -i \gamma_0 \gamma_i \partial^i \psi_-
+ m\gamma_0 \psi_- + g \gamma_0 \phi   \psi_-        $$
$$ i \partial^- \psi_- = -i \gamma_0 \gamma_i \partial^i \psi_+
+ m\gamma_0 \psi_+ + g \gamma_0 \phi   \psi_+ \eqn\b $$
where $\psi_\pm = \Lambda_\pm \psi$ and $\Lambda_\pm = {1\over 2}\gamma^0
\gamma^\pm$  . In the usual front-form quantization, $\psi_+$ is taken
to be the independent degree of freedom and $\psi_-$ the {\it dependent}
degree of freedom.

Let us look now at the scalar field. The equations of motions for it are
$$    \partial^+ \partial^- \phi-  \partial_i ^2 \phi +
\mu^2 \phi = - g \overline{\psi} \psi \eqn\c$$
and we want to point out that the fermionic current term is of the
form
$$ \overline{\psi} \psi = \psi^{\dagger}_+ \gamma_0 \psi_- +
\psi^{\dagger}_- \gamma_0 \psi_+  \eqn\d$$

Robertson and McCartor invite us to look at the evolution of $\phi$
along $x^+$
$$\partial_- \phi_{x^+ = \delta x^+ }
= \partial_-\phi_{x^+ = 0} + (\delta x^+ ) {1\over 4} (\partial_i ^2
 \phi - \mu^2 \phi - g\overline{\psi}\psi  )_{x^+ = 0}
\eqn\e$$
As they point out , as we follow the evolution of $\phi$, in order
 to determine $\phi$ everywhere we need
to know $\phi$ along {\bf both} $x^+ = 0$ and $x^- = 0$ surfaces
\REF\ch{
Another argument is as follows: In equal-time quantization,
one specifies the initial data along a space-like surface $t=0$    .
Evolution of that system corresponds to specifying the system at other
surfaces $t=c_1 $, $t=c_2 $ and so on. Now for the type of equation
we are considering (hyperbolic, second order), the characteristic
curve lies along the  surfaces $x^+=0$ and $x^- =0$. For
$t\ne 0$, our equal-time
quantization surface {\bf will} intersect both
characteristic surfaces, hence the solution will be well defined.
Now, when we consider quantization along $x^+ =0 $, (one of the
characteristic surfaces) evolving
       the solution along $x^- $ for different constant $x^+$'s
    will {\bf never} intersect the   characteristic surface
$x^+ = 0$ . This  seems to say that we will not have a consistent
solution to our equations.
Therefore,  if we intend  to quantize our theory along
 characteristic surfaces, we {\bf must} take all  these
surfaces  as our quantization surfaces.
In our case, this means that we need to use both $x^+  = 0$ and  $x^- =0$
as quantization surfaces on the front-form.
}
\refend .

We'd like to point out that in some work involving initial value problems
in gravity using front-form coordinates
\REF\ivpa{R. Gambini, A. Restuccia, Phys. Rev. \us{D17},
3150, (1978)}
\refend
\REF\ivpb{R. K. Sachs, J. Math. Phys. \us{3}, 908, (1962)}
\refend
\REF\ivpc{R. Penrose, Gen. Relat. and Grav. \us{12}, 225, (1980);
Roger Penrose and Wolfgang Rindler, 'Spinors and Space-Time,' vol. I,
Chapter 5 ,Cambridge University Press, New York, 1984 }
\refend
\REF\ivpd{H. M\"uller zum Hagen and H-J Seifert,
Gen. Relat. and Grav. \us{ 8}, 259, (1977)}
\refend
\REF\ivpe{H. Bondi, M. G. J. van der Burg and A. W. K. Metzner,
Proc. R. Soc. London, \us{A269}, 21, (1962)}
\refend  ,
the initial data for these coordinates is
also specified along {\bf both} $x^+ = const$ and $x^- = const$
surfaces as well as at $x^+ =x^- = 0$ .
Furthermore, R. Penrose [\ivpc]
points out that in this approach there are {\it no} constraints
\REF\noconst{ Most of the work on front-form quantization is motivated
by the desire to study (maybe 'solve') $QCD$ .
The  lack of constraints could have repercussions far
beyond  front-form quantization of $QCD$ ! There is much work
[\cd], [\gt], [\gov], [\ht],
dealing with quantizing systems with constraints. The  present
approach could
bypass much of the difficulty coming from the presence of constraints.
It is not clear to us yet whether the lack of constraints holds
in the presence of gauge symmetries , local or global . }
\refend   .
\chapter{Reduced Phase Space Quantization in Front-Form}
We will apply the reduced phase space quantization of Faddeev and
Jackiw
\REF\fj{L. D. Faddeev and R. Jackiw,  Phys. Rev. Lett. \us{60},
1692,(1988)}
\refend  .
to the Yukawa model in $3+1$ dimensions with boson
mass $\mu$ and fermion mass $m$:
$$\L = {1\over2}(\partial_\nu \phi \partial^\nu \phi - \mu^2 \phi^2)
+ {i\over 2} \partial^{\nu}\overline{\psi} \gamma_{\nu} \psi
- \overline{\psi} {i\over 2} \partial^{\nu}\gamma_{\nu} \psi
 -m \overline{\psi} \psi  + \L_I \eqn\f $$
where $\L_I $ is the interaction part
of the Lagrangean. Let us write  $\L$ out explicitly :
$$\L d^4 x = \big\{ {1\over 2} \partial^+ \phi \partial^- \phi -
{1\over 2}(\partial_i \phi)^2 -{1\over 2}\mu^2 \phi^2 +\L_I +
\psi^{\dagger}_+ {i\partial^- \over 2}\psi_+ - {i\over 2}
(\partial^- \psi^{\dagger}_+ )\psi_+   $$
$$ +\psi^{\dagger}_- {i\partial^+ \over 2}\psi_- - {i\over 2}
(\partial^+ \psi^{\dagger}_- )\psi_- +
-\psi^{\dagger}_+\gamma_0\gamma_i {i\partial_i \over 2}\psi_- +
{i\over 2}(\partial_i\psi^\dagger_-)\gamma_0\gamma_i\psi_+  $$
$$ -\psi^{\dagger}_-\gamma_0\gamma_i {i\partial_i \over 2}\psi_+ +
{i\over 2}(\partial_i\psi^\dagger_+)\gamma_0\gamma_i\psi_- -
m\psi^{\dagger}_+{\gamma^- \over 2}\psi_- -
m\psi^{\dagger}_-{\gamma^+ \over 2}\psi_+ \big\}
{dx^- dx^+ d^2 x_\perp \over 2}    \eqn\fg$$
where $\psi_\pm = \Lambda_\pm \psi$ as before. Note also that
$\partial_\nu = {\partial \over \partial x^\nu}$ so that
$\partial^- = 2 \partial_ + = 2 {\partial \over \partial x^+ }$ and
$\partial^+ = 2 \partial_ - = 2 {\partial \over \partial x^- }$ .
The corresponding  conjugate momenta for     $x^+$-derivatives are
$$ \pi_\phi =  {   \partial \L \over \partial (\partial^{-} \phi   ) } =
 {1\over 2}\partial^+ \phi \eqn\h $$
$$\pi_{\psi}(x)= {\partial \L \over \partial (\partial^{-} \psi) } =
 {i\over 2}\psi^{\dagger}_{+} \eqn\i$$
$$\pi_{\psi^{\dagger}}(x)={\partial \L \over
  \partial (\partial^{-} \psi^{\dagger}) }=-{i\over 2}\psi_{+} \eqn\j$$
For the momenta corresponding to $x^-$-derivatives we get similar forms:
$$\rho_\phi =  {   \partial \L \over \partial (\partial^{+} \phi   ) } =
 {1\over 2}\partial^- \phi \eqn\hh$$
$$\rho_{\psi}(x)={\partial \L \over  \partial (\partial^{+} \psi) } =
 {i\over 2}\psi^{\dagger}_{-} \eqn\k$$
$$\rho_{\psi^{\dagger}}(x)=
{\partial \L \over \partial (\partial^{+} \psi^{\dagger}) }=
-{i\over2}\psi_{-} \eqn\l$$
We rewrite $\L d^4 x$ in the following way
$$\L d^4 x = {1\over 2}\big\{ \pi_\phi \partial^-\phi+\pi_\phi
\partial^-\phi +\pi_\psi\partial^-\psi_+  +\pi_\psi\partial^-\psi_+  +
(\partial^-\psi^{\dagger}_+)\pi_{\psi^{\dagger}}  $$
$$ +(\partial^-\psi^{\dagger}_+)\pi_{\psi^{\dagger}} +
\rho_\phi\partial^+\phi + \rho_\phi\partial^+\phi +
    \rho_\psi\partial^+\psi_- +\rho_\psi\partial^+\psi_-  +
(\partial^-\psi^{\dagger}_-) \rho_{\psi^{\dagger}} +
(\partial^-\psi^{\dagger}_-) \rho_{\psi^{\dagger}} \big\}d^4 x $$
$$ -\H dx^+  -\K dx^- +\M d^4 x \eqn\fjj$$
The meaning of these terms is as follows : the first bracket represents
the p-q-dot terms which go into the definitions of the canonical
commutation relations; the second and third term are the Hamiltonians
which  define the evolution of the system along $x^+$ , given by $\H$,
and along $x^-$ given by  $\K$
\REF\hv{ Note that the dynamical nature of $P^+$, corresponding to
$\K$  ( $P^-$ corresponds to $\H$), was noticed already
by A. Harindranath and J. Vary in their study of $\phi^4$ in
$1+1$-dimensions [ Phys. Rev. {\us D 37}, 3010 (1988) ] . They
found that in front-form quantization, $P^+$ is also affected by
operator ordering, in addition to the  Hamiltonian $P^-$. The corrections
in $P^+$ due to operator ordering amount to making it dynamic. The
mystery is why is that so, since $P^+$ is supposed to be kinematic.
An attractive
feature of the present work is that $P^+$ becomes dynamical naturally.}
\refend
; finally, the last term contains
the remaining
pieces which give the 'constraints', though are not  'true constraints'
[\fj] as are consistent with the $\phi$ equations of motions. The
Hamiltonians $\H$ and $\K$ are :
$$\H = {dx^- dx^2_\perp \over 2}\big\{ {1\over 2}(\partial_i\phi)^2 +
{1 \over 2}\mu^2 \phi^2 -\L_I
+\psi^{\dagger}_+\gamma_0\gamma_i {i\partial_i \over 2}\psi_- -
{i\over 2}(\partial_i\psi^\dagger_-)\gamma_0\gamma_i\psi_+  $$
$$ +m\psi^{\dagger}_+{\gamma^- \over 4}\psi_- +
m\psi^{\dagger}_-{\gamma^+ \over 4}\psi_+\big \} \eqn\pmin$$
$$\K = {dx^+ dx^2_\perp \over 2}\big\{ {1\over 2}(\partial_i\phi)^2 +
{1 \over 2}\mu^2 \phi^2 -\L_I
+\psi^{\dagger}_-\gamma_0\gamma_i {i\partial_i \over 2}\psi_+ -
{i\over 2}(\partial_i\psi^\dagger_+)\gamma_0\gamma_i\psi_- $$
$$ +m\psi^{\dagger}_-{\gamma^+ \over 4}\psi_+ +
m\psi^{\dagger}_+{\gamma^- \over 4}\psi_-\big \} \eqn\pplu$$
and for the 'constraints' we get
$$\M = \big\{-{1\over 2} \partial^+ \phi \partial^- \phi +
{1\over 2}(\partial_i \phi)^2 +{1\over 2}\mu^2 \phi^2 -\L_I
\big\} \eqn\mt$$

In what sense is the last term $\M$ not a constraint ? Well, we can
rewrite is as
$$ \M d^4 x = {1\over 2}\M d^4 x+ {1\over 2}\N d^4 x \eqn\mmn $$
with $\M =\N$ . Now we rewrite each of these terms thus
$$\M = \int \delta \M  ,\quad \N = \int \delta \N  \eqn\varmn$$
where the variations $\delta$ are all possible variations over the
field $\phi$ . Using the usual Euler-Lagrange equations, this gives
(up to total derivatives which we can discard):
$$\M d^4 x = {1\over 2}\int \delta\phi \big\{ \partial^\nu {\partial
\M \over \partial(\partial^\nu \phi)}-{\partial \M \over \partial
\phi} \big\} d^4x +
 {1\over 2}\int \delta\phi \big\{ \partial^\nu {\partial
\N \over \partial(\partial^\nu \phi)}-{\partial \N \over \partial
\phi} \big\} d^4 x\eqn\elvarmn $$
For the form $\M$, this gives
$$ \M d^4x = {1\over 2}\pi_\phi C_\pi d^4x +
 {1\over 2}\rho_\phi C_\rho d^4 x \eqn\notconst $$
where we used the definitions
$$\delta \phi = \pi_\phi \delta x^+ = \rho_\phi \delta x^-
\eqn\defvar $$
and the 'constraints' $C_\pi$ and $C_\rho$ are
$$C_\pi = \int \delta x^+ \big\{ -\partial^+
\partial^- \phi + \partial_i^2\phi - \mu^2 \phi +{\partial\L_I \over
\partial \phi} \big\} \eqn\cpi $$
$$C_\rho =  \int \delta x^- \big\{ -\partial^+
\partial^- \phi + \partial_i^2\phi - \mu^2 \phi +{\partial\L_I \over
\partial \phi} \big\} \eqn\crho $$
We see that $C_\pi = C_\rho = 0$ identically by the equation    of
motion, so in that sense these are not new conditions, so are not 'true
constraints' [\fj] .

Let us write the $\L dx^4$ with the explicit momenta dependence ( up
to total derivatives which we can discard [\fj],
\REF\zh{Wei-Min Zhang and Avaroth Harindranath ,'Light-Front QCD:
Role of Longitudinal Boundary Integrals', hepth@xxx/9302119;
Wei-Min Zhang and Avaroth Harindranath ,'Residual Gauge Fixing in
Light-Front QCD', hepth@xxx/9302107}
\refend   ),
so as to make the resulting commutation relation clear :
$$\L d^4 x = {1\over 2}2\big\{ \pi_\phi d \phi- \phi d \pi -\phi +
\pi_\psi d\psi_+  -d \pi_\psi    \psi_+  +d\psi^{\dagger}_+
\pi_{\psi^{\dagger}} - \psi^{\dagger}_+ d \pi_{\psi^{\dagger}}
\big\} {dx^-dx_\perp  \over 2}   $$
 $$ +        {1\over 2}2\big\{\rho_\phi d\phi- \phi d \rho_\phi +
 \rho _\psi d \psi_- -d \rho _\psi \psi_- +d\psi^{\dagger}_-
 \rho _{\psi^{\dagger}} - \psi^{\dagger}_- d\rho_{\psi^{\dagger}}
 \big\} {dx^+dx_\perp \over 2}  $$
$$ -\H dx^+  -\K dx^- + {1\over 2}\pi_\phi C_\pi d^4x +
 {1\over 2}\rho_\phi C_\rho d^4 x \eqn\cc $$
We see now that we have two types of evolutions, one along $x^+$, for
which the first term in equation  \cc\quad gives the commutation
relations along surfaces $x^+ = y^+ $ according to the form:
$$ [\xi^a , \xi^b ] = \Gamma ^{-1} _{ab}\quad a,b=1,..6\eqn\xplucc $$
with
$$\xi^1 = \pi_\phi , \xi^2 = \pi_\psi, \xi^3 = \pi_{\psi^{\dagger}} ,
\xi^4 = \phi, \xi^5 = \psi_+, \xi^6 = \psi_+ ^{\dagger}\eqn\xis $$
and
$$\Gamma_{14} = \Gamma_{25} = \Gamma_{36} = 2 =
-\Gamma_{41} = -\Gamma_{52} =  -\Gamma_{63}  \eqn\gs$$
and all the other $\Gamma$'s are $0$ .
The second term in equation  \cc\quad gives the commutation
relations along surfaces $x^- = y^- $ according to the form :
$$ [\eta ^a ,\eta^b ] = \Delta ^{-1} _{ab}\quad a,b=1,..6\eqn\xmincc $$
with
$$\eta^1 = \rho_\phi, \eta^2 = \rho_\psi, \eta^3 =\rho_{\psi^{\dagger}} ,
\eta^4 = \phi, \eta^5 = \psi_-, \eta^6 = \psi_- ^{\dagger}\eqn\eis $$
and
$$\Delta_{14} = \Delta_{25} = \Delta_{36} = 2 =
-\Delta_{41} = -\Delta_{52} =  -\Delta_{63}  \eqn\ds$$
and all the other $\Delta$'s are $0$ .
Going now to the quantum commutators, we get the following relations
for fields at equal $x^+ = y^+ $ , the usual front-form 'time' :
$$ [\phi(x^+ , x^- ,x_\perp) , \pi_\phi (y^+, y^-, y_\perp) ] _{x^+ =
y^+} = {i \over 2}\delta(x^- - y^-)\delta^2(x_\perp - y_\perp)
\eqn\pfxplu$$
$$\big\{\psi_+(x^+,x^-,x_\perp),\pi_\psi (y^+,y^-,y_\perp)\big\}_
{x^+ = y^+} = +{i\over 2} \Lambda_+ \delta(x^- - y^-)
\delta^2(x_\perp - y_\perp) \eqn\psii$$
$$\big\{\psi_+^{\dagger}(x^+,x^-,x_\perp),
\pi_{\psi^{\dagger}} (y^+,y^-,y_\perp)\big\}_
{x^+ = y^+} = -{i\over 2} \Lambda_+ \delta(x^- - y^-)
\delta^2(x_\perp - y_\perp) \eqn\psidag $$

For fields at equal $x^- = y^-$, a new front-form 'time', we get:
$$ [\phi(x^+ , x^- ,x_\perp) ,\rho_ \phi (y^+, y^-, y_\perp) ] _{x^- =
y^-} = {i \over 2}\delta(x^- - y^-)\delta^2(x_\perp - y_\perp)
\eqn\pfxmin$$
$$\big\{\psi_-(x^+,x^-,x_\perp),\rho_\psi (y^+,y^-,y_\perp)\big\}_
{x^- = y^-} = +{i\over 2} \Lambda_- \delta(x^+ - y^+)
\delta^2(x_\perp - y_\perp) \eqn\psim $$
$$\big\{\psi_-^{\dagger}(x^+,x^-,x_\perp),
\rho_{\psi^{\dagger}} (y^+,y^-,y_\perp)\big\}_
{x^- = y^-} = -{i\over 2} \Lambda_- \delta(x^+ - y^+)
\delta^2(x_\perp - y_\perp) \eqn\psimdag $$

The equations of motions are now  like in Faddeev and Jackiw [\fj]
$$ \Gamma _{ab} \partial^- \xi^b = {\partial \H \over \partial \xi^a}
\eqn\eomh $$
for the $x^+$ variation, and
$$ \Delta _{ab} \partial^- \eta^b = {\partial \K \over \partial
\eta^a}\eqn\eomk $$
for the $x^-$ variation .
For $a=4$ and $b=1$, equation \eomh\quad gives
$$ \partial^+\partial^- \phi =
 -\partial_i^2\phi + \mu^2 \phi -{\partial\L_I \over
\partial \phi}  \eqn\eomf$$
For $a=5$ and $b=2$ we recover the equation of motion for
$\psi_+^{\dagger}$
$$ i \partial ^- \psi_+^{\dagger} = i {\partial_i \psi_-^{\dagger}
\over 2}\gamma_0\gamma_i - m\psi_-^{\dagger}{\gamma_0 \over 2}
-{\partial \L_I \over \partial \psi_+ } \eqn\eompsid $$
We get similar results from \eomk . Note that for $a=1$
and $b=4$ we get a seeming contradiction :
$$2 \partial^- \phi = 4 \pi_\phi = 0 , 2 \partial^+ \phi =
4 \rho_\phi = 0 \eqn\phiconst $$
But this is just why we have the $\M$ term, which contains the
'constraints' of the theory [\fj], [\zh] :
it is of the form \notconst\quad , where the $\pi_\phi$ and
$\rho_\phi$ are the Lagrange multipliers and the $C$'s are the
'constraints'.

\chapter{Quantization of the Fields}
Now that we have the commutation relations, we are ready to define
the fields $\phi$ and $\psi$.  According to [\ivpc ], using two null
hyperplanes, the initial data must be specified on each of the
hyperplanes as well as on their intersection .
In this case,    we will have  initialization
 on the two surfaces $x^+ =0$ and $x^- =0$ . We will
require, though, that on the intersection of these surfaces, at
$x^+ =x^- = 0$ these fields satisfy certain consistency conditions.
This works out as follows.

On $x^+ =0$ we have then :
$$ \phi(x^+ = 0, x^- , x_\perp ) = \int {d^2 k_\perp \over (2\pi)^3 }
{dk^+ \over 2 k^+ } \big\{ a(k^+ , k_\perp ) e^{-i k . x} +
a^{\dagger}(k^+ , k_\perp ) e^{+i k . x} \big\} \eqn\phipl $$
$$\psi_+(x^+ = 0, x^- , x_\perp) = \int {d^2 k_\perp \over (2\pi)^3 }
{dk^+ \over 2 k^+ }\sum_{\lambda} \big\{ b(k^+ , k_\perp ) u_+(k^+ ,
k_\perp, \lambda) e^{-i k . x} $$
$$ + d^{\dagger}(k^+ , k_\perp)
v_+(k^+ , k_\perp, \lambda) e^{+i k . x} \big\} \eqn\psipl $$
In this case, $i k . x = i k^+ x^- - i k_\perp . x_\perp $ .

On the other hyperplane, $x^- = 0$ we get similar forms:
$$ \phi(x^- = 0, x^+ , x_\perp ) = \int {d^2 k_\perp \over (2\pi)^3 }
{dk^- \over 2 k^- } \big\{ \hat a(k^- , k_\perp ) e^{-i \hat k . x} +
\hat a^{\dagger}(k^- , k_\perp ) e^{+i \hat k . x} \big\} \eqn\phimi $$
$$\psi_-(x^- = 0, x^+ , x_\perp) = \int {d^2 k_\perp \over (2\pi)^3 }
{dk^- \over 2 k^- }\sum_{\mu} \big\{ \hat b(k^- , k_\perp ) u_-(k^- ,
k_\perp, \mu) e^{-i \hat k . x} $$
$$ + \hat d^{\dagger}(k^- , k_\perp)
v_-(k^- , k_\perp, \mu) e^{+i \hat k . x} \big\} \eqn\psimi $$
Here  , $i \hat k . x = i k^- x^+ - i k_\perp . x_\perp $ .

We require now that the fields  be consistent at $x^+ = x^- = 0$. This
means  that we have
$$ \phi (x^+ = 0, x^- = 0, x_\perp) = \phi (x^- = 0, x^+ = 0, x_\perp)
\eqn\phiphi $$
a tautology, obviously true. This implies
$$  \int {d^2 k_\perp \over (2\pi)^3 }
{dk^+ \over 2 k^+ } \big\{ a(k^+ , k_\perp ) e^{+i k_\perp . x_\perp} +
a^{\dagger}(k^+ , k_\perp ) e^{-i k_\perp . x_\perp} \big\} $$
$$  = \int {d^2 k_\perp \over (2\pi)^3 }{dk^- \over 2 k^- }
\big\{ \hat a(k^- , k_\perp ) e^{+i k_\perp . x_\perp} + \hat a^{\dagger}
(k^- , k_\perp ) e^{-i k_\perp . x_\perp}\big\}\eqn\aadag $$
As $k^+$ and $k^-$ are just dummy variables here, we get that
$$a(k^+ , k_\perp) = \hat a(k^+  , k_\perp),\quad a^{\dagger}
(k^+ , k_\perp) = \hat a^{\dagger}(k^+  , k_\perp) \eqn\aad$$
and we need to point out that the  variables are the {\bf same} for
both creation operators.
So this means that
$$a(k^+ , k_\perp) \ne \hat a(k^- , k_\perp) \eqn\aane$$
hence the field  $\phi$ has different effects on the two surfaces.
On $x^+ =0$, $\phi(x^+ = 0, x^- , x_\perp)$
creates or destroys scalar quanta with momentum
$ k = (k^+ , k_\perp) $ . On $x^- =0$, $\phi(x^- =0, x^+ , x_\perp)$
creates or destroys quanta with momentum $\hat k = (k^- , k_\perp)$.

A similar though more involved analysis goes for the fermion fields.
Equating $\psi_+(x^+ =0, x^- , x_\perp)$
and $\psi_-(x^- =0, x^+ , x_\perp)$ at $x^+ = x^- = 0$ we get
$$\psi_+(x^+ =0, x^- =0, x_\perp) = \psi_-(x^- =0, x^+ =0, x_\perp)
\eqn\psps $$
which gives
$$ \int {d^2 k_\perp \over (2\pi)^3 }
{dk^+ \over 2 k^+ }\sum_{\lambda} \big\{ b(k^+ , k_\perp ) u_+(k^+ ,
k_\perp, \lambda) e^{+i k_\perp . x_\perp} $$
$$ + d^{\dagger}(k^+ , k_\perp)
v_+(k^+ , k_\perp, \lambda) e^{-i k_\perp . x_\perp} \big\} $$
$$  = \int {d^2 k_\perp \over (2\pi)^3 }
{dk^- \over 2 k^- }\sum_{\mu} \big\{ b(k^- , k_\perp ) u_-(k^- ,
k_\perp, \mu) e^{+i k_\perp . x_\perp} $$
$$ + d^{\dagger}(k^- , k_\perp)
v_-(k^- , k_\perp, \mu) e^{-i k_\perp . x_\perp} \big\} \eqn\pspsm$$
After some substitutions,
this gives    the following relations
for the fermion fields and the associate spinors
$$b(k^+ , k_\perp) = \hat  b(k^+ , k_\perp) , \quad
d^{\dagger}(k^+ , k_\perp) = \hat  d^{\dagger}
(k^+ , k_\perp) \eqn\psipsi $$
$$u_+ (k^+ , k_\perp) =u_- (k^+ , k_\perp)=u(k^+ , k_\perp) \eqn\usp$$
$$v_+ (k^+ , k_\perp) =v_- (k^+ , k_\perp)=v(k^+ , k_\perp) \eqn\vsp$$
Note again that the variables for the creation, destruction operators as
well as for the spinors are  all the {\bf same} . On the other hand,
$\psi_+(x^+ =0, x^- , x_\perp)$ and $\psi_-(x^- =0, x^+ , x_\perp)$
act differently on the two surfaces: on $x^+ =0$, $\psi_+(x^+ =0,x^- ,
x_\perp)$ creates
or destroys fermion quanta of momentum
$ k = (k^+ , k_\perp) $ . On $x^- =0$,   $\psi_-(x^- =0, x^+ ,
x_\perp)$
 creates or destroys quanta with momentum $\hat k = (k^- , k_\perp) $ .

Let us also point out that these relationships between $a$ and $\hat a$,
$b$ and $\hat b$, $d$ and $\hat d$, $u_+$ and $u_-$, $v_+$ and $v_-$
guarantee  that we have no doubling of the independent degrees of freedom
, a possibility due to the presence of {\bf two} initializing surfaces:
we have the same number of creation, destruction operators and of spinors
as in the equal-time quantization case.

\chapter{Parity in Front-Form Quantization}
We are ready now to study how the fields $\phi$  ,  $\psi_+$ and
$\psi_-$ transform under parity. For this we use the usual definition
(Bjorken and Drell for instance
\REF\bd{James  D. Bjorken and Sidney D. Drell, 'Relativistic Quantum
Fields,' Chapter 15 , McGraw-Hill, San Francisco, 1965}
\refend  ) :
$$ \P \phi(x^+ , x^- , x_\perp) \P^{-1} = \pm \phi(x^- , x^+ , -x_\perp)
\eqn\scpar $$
since under parity $(x^+ , x^- , x_\perp) \rightarrow (x^- , x^+ ,
-x_\perp)$ . The $\pm$ in front of the scalar field represent the
intrinsic parity of the field.
For the scalar field we get
$$ \P\phi(x^+ = 0, x^- , x_\perp )\P^{-1} =\P
\int {d^2 k_\perp \over (2\pi)^3 }
{dk^+ \over 2 k^+ } \big\{ a(k^+ , k_\perp ) e^{-i k . x} +
a^{\dagger}(k^+ , k_\perp ) e^{+i k . x} \big\}\P^{-1} \eqn\phipa $$
This becomes
$$ \P\phi(x^+ = 0, x^- , x_\perp )\P^{-1} =
   \int {d^2 (-k_\perp) \over (2\pi)^3 }
{dk^- \over 2 k^- } \big\{ a(k^- ,-k_\perp ) e^{-i k'. x'} +
a^{\dagger}(k^- ,-k_\perp ) e^{+i k'. x'} \big\}\eqn\phipb $$
if
$$\P a(k^+ ,k_\perp) \P^{-1} = a(k^- , -k_\perp),\quad \P a^{\dagger}
(k^+ ,k_\perp) \P^{-1} = a^{\dagger}(k^- , -k_\perp) \eqn\apa $$
and $i k' . x' = ik^- x^+ - i k_\perp x_\perp$ . Redefining variables
$(k^-, -k_\perp)\rightarrow (l^-,l_\perp)$, we get the result
$$ \P \phi(x^+ =0, x^- , x_\perp) \P^{-1} = \phi(x^- =0, x^+ , -x_\perp)
  \eqn\phipc$$
Let us consider the fermion fields now. In this case we have [\bd]
$$ \P \psi (x^+ , x^- , x_\perp) \P^{-1} = \gamma_0 \psi (x^- , x^+ ,
-x_\perp) \eqn\psipa $$
and   we expect that fields defined on $x^+$ will be mapped into
fields defined on $x^-$ by parity. Indeed, that is what we find for
$\psi_+$ :
$$\P \psi_+ (x^+ = 0, x^- , x_\perp) \P^{-1} = \P
         \int {d^2 k_\perp \over (2\pi)^3 }
{dk^+ \over 2 k^+ }\sum_{\lambda} \big\{ b(k^+ , k_\perp ) u(k^+ ,
k_\perp, \lambda) e^{-i k . x} $$
$$ + d^{\dagger}(k^+ , k_\perp)v(k^+ , k_\perp, \lambda)
e^{+i k . x} \big\}\P^{-1} \eqn\psipb $$
This becomes
$$\P \psi_+ (x^+ = 0, x^- , x_\perp) \P^{-1} =
            \int {d^2(- k_\perp) \over (2\pi)^3 }
{dk^- \over 2 k^- }\sum_{\lambda} \big\{ b(k^- ,-k_\perp ) u(k^- ,-
k_\perp, \lambda) e^{-i k'. x'} $$
$$ + d^{\dagger}(k^- ,-k_\perp)v(k^- ,-k_\perp, \lambda)
e^{+i k'. x'} \big\} \eqn\psipc $$
if creation, destroying operators transform under parity thus :
$$\P b(k^+ , k_\perp) \P^{-1} = b(k^- , -k_\perp) ,\quad
\P d^{\dagger}(k^+ , k_\perp) \P^{-1} =
 -d^{\dagger}(k^- , -k_\perp)  \eqn\psipd $$
 and if the spinors transform thus :
$$\gamma_0 u(k^+ , k_\perp) = + u(k^- , -k_\perp) , \quad
  \gamma_0 v(k^+ , k_\perp) = - v(k^- , -k_\perp) \eqn\psipe $$
A rather long and subtle  but essentially straightforward calculation
show that this is indeed the case
\REF\dm{ Mikolaj Sawicki and Dharam Ahluwalia, ' Spinors on the
Front-Form,' LANL preprint, October 1993}
\refend .
After manipulations similar to  the
scalar case and after using the fact that
    $i k' . x' = ik^- x^+ - i k_\perp x_\perp$ as well as
    redefining variables  , we obtain the following result :
$$ \P \psi_+(x^+ =0, x^- , x_\perp) \P^{-1} = \gamma_0 \psi_-
(x^- =0, x^+ , -x_\perp) \eqn\psipf $$

We derive now these relations for arbitrary $x^+$ and $x^-$.
Note that for the $x^+$  evolution  we have
$$ \phi(x^+ , x^- , x_\perp)  = e^{-iP^- x^+} \phi
(x^+ =0 , x^- , -x_\perp) \eqn\evolphi $$
or
$$ \psi_- (x^+ , x^- , x_\perp)  = e^{-iP^- x^+} \psi_-
(x^+ =0 , x^- , -x_\perp) \eqn\evolphi $$
so that the parity-transformed field is
$$ \P\phi(x^+ , x^- , x_\perp)\P^{-1}  = \P e^{-iP^- x^+} \P^{-1}
\P \phi(x^+ =0 , x^- , -x_\perp)\P^{-1} \eqn\evolphib$$
which becomes
$$ \P \phi(x^+ , x^- , x_\perp)\P^{-1}  = e^{-iP^+ x^-} \phi
(x^- =0 , x^+ , -x_\perp) \eqn\evolphic$$
since
$$\P P^{-} \P^{-1} = \P \int \H \P^{-1} = \int \K = P^{+} \eqn\parh$$
by use of the equations \pmin\quad and \pplu.
A similar result holds for the fermion case.
We also get the generator  of $x^-$  evolutions to transform properly
as well since
$$\P P^{+} \P^{-1} = \P \int \K \P^{-1} = \int \H = P^{-} \eqn\park$$
again, by use of equations \pplu\quad and \pmin.

Since now the generators of evolution along $x^+$ and $x^-$ ($\H$ and
$\K$ respectively), transform properly under parity ,
we can evolve the
parity relations obtained at $x^+ =0$ and $x^- =0$ to
relations for arbitrary $x^+$ and $x^-$. For the scaler case we get
$$ \P \phi(x^+ , x^- , x_\perp) \P^{-1} = \phi(x^- , x^+ , -x_\perp)
  \eqn\phipd$$
as expected from previous work [\pclcq] .

For the fermion  case, we get
$$ \P \psi_+(x^+ , x^- , x_\perp) \P^{-1} = \gamma_0 \psi_-
(x^- , x^+ , -x_\perp) \eqn\psiph $$
which show very clearly that parity maps independent fields on
$x^+ =0$ [$\psi_+(x^+ =0, x^- , x_\perp)$] ,
to independent fields on $x^- =0$  [$\psi_-(x^- =0, x^+ , x_\perp) $] ,
demonstrating the it is crucial that
we take {\bf both} $x^+ =0$ and $x^- =0$ as quantizing surfaces if we
desire to have fields with parity as an explicit symmetry.

Thus far we have looked at transformation properties of independent
fields on $x^+ =0$ . It is quite straightforward to show that we get
similar results for the fields which are initialized on $x^- =0$ :
$$ \P \phi(x^- , x^+ , x_\perp) \P^{-1} = \phi(x^+ , x^- , -x_\perp)
  \eqn\phipe$$
for the scalar field  and
$$ \P \psi_-(x^- , x^+ , x_\perp) \P^{-1} = \gamma_0 \psi_+
(x^+ , x^- , -x_\perp) \eqn\psipg $$
for the fermion field. This completes our demonstration that fields
defined on $x^+ =0$ and $x^- =0$ transform properly under parity, and
are defined consistently.

\chapter{Acknowledgements}

I  would like to thank Prof. Stanley Brodsky and Dr. Dharam Ahluwalia
for discussions and to thank Prof. Richard Blankenbecler
for his continuing support; I  would also like to thank the organizers
of the Dallas Workshop on Light-Cone Quantization in May 1992
for putting together
such a simulating meeting which gave impetus to this work.
\endpage
\refout
\end